\documentclass[12pt]{article}
\usepackage{times}
\usepackage{graphicx}
\usepackage{multirow}
\usepackage{color}
\usepackage{bm}

\title{Mapping partial wave dynamics in scattering resonances by rotational de-excitation collisions}

\author
{Tim de Jongh$^1$ $^{\ast \dagger}$, Quan Shuai$^1$ $^{\ast\dagger\dagger}$, Grite L. Abma$^{1}$, Stach Kuijpers$^{1}$, Matthieu Besemer$^1$,\\ Ad van der Avoird$^1$, Gerrit C. Groenenboom$^1$,$^{\ast\ast}$ \\ Sebastiaan Y.T. van de Meerakker$^1$ $^{\ast\ast}$ \\  \\
\\
\normalsize{$^1$Radboud University, Institute for Molecules and Materials}\\
\normalsize{Heyendaalseweg 135, 6525 AJ Nijmegen, the Netherlands}\\
\normalsize{$^{\dagger}$ Present address: Laboratoire Kastler Brossel, ENS-Universit{\'e} PSL,} \\
\normalsize{CNRS, Sorbonne Université, Collège de France, 24 rue Lhomond, 75005 Paris, France}\\
\normalsize{$^{\dagger\dagger}$ Present address: State Key Laboratory of Molecular
Reaction Dynamics,}\\
\normalsize{Dalian Institute of Chemical Physics, Chinese Academy of Sciences, Dalian, 116023, China.}\\
\normalsize{$^{\ast}$ Who contributed equally to this work}\\
\normalsize{$^{\ast\ast}$ To whom correspondence should be addressed}\\
\normalsize{E-mail: basvdm@science.ru.nl, gerritg@theochem.ru.nl}}

\begin{document}
\date{\today}

\maketitle

\begin{abstract}
One of the most important parameters in a
collision is the `miss distance' or impact parameter, which in quantum
mechanics is described by quantized partial waves. Usually, the
collision outcome is the result of unavoidable averaging over many
partial waves. Here we present a study of low-energy NO\textendash He collisions, that enables us to probe how
individual partial waves evolve during the collision. By tuning the
collision energies to scattering resonances between 0.4 and 6 cm$^{-1}$,
the initial conditions are characterized by a limited set of partial
waves. By preparing NO in a rotationally excited state before the
collision and by studying rotational de-excitation collisions, we were
able to add one quantum of angular momentum to the system and trace how
it evolves. Distinct fingerprints in the differential cross sections
yield a comprehensive picture of the partial wave dynamics during the
scattering process. Exploiting the principle of detailed balance, we
show that rotational de-excitation collisions probe time-reversed
excitation processes with superior energy and angular resolution.

\end{abstract}

\newpage
Understanding the interactions between individual atoms and molecules at
the complete quantum level is of fundamental importance to physics and
chemistry. Collision experiments have proven an indispensable tool to
unravel these interactions. The key to understanding a collision lies
with investigating how the interaction potential transforms the reagent
states into the product states.
Experimentally, this is best probed by so-called state-to-state
measurements, in which the initial conditions are precisely controlled
and the final conditions are precisely detected. To date, state-to-state
measurements for (almost) all degrees of freedom have become possible.
Ingenious methods have for instance been developed to prepare a single
rotational or vibrational state of a molecule before the collision, and
to detect it state-selectively afterwards
\cite{Chadwick:JCP140:034321,Perreault:science358:356,McDonald:Nature535:122,Hu2019}.

In addition to the quantized energy levels of an isolated molecule,
quantization also appears when the molecule interacts with another
particle. The relative motion of the particles is perceptibly quantized
in terms of the orbital angular momentum, labeled by the integer $\ell$,
which replaces the classical notion of the impact parameter by its
quantum mechanical analogue of partial waves. The nomenclature of these
discrete orbital states is similar to that of the electronic states of
the hydrogen atom: $\ell = 0, 1, 2, ...$ are named $s$, $p$, $d$, ...
waves, respectively.

Gaining control over this relative motion (i.e., impact parameter or
partial waves) is a long-standing dream in state-to-state measurements
\cite{Herschbach1987}. Indeed, the ultimate information about a
collision event would be found by selecting a single partial wave
$\ell_\mathrm{in}$ from the initial conditions, and by subsequently
following how the interaction transforms it into product partial waves
$\ell_\mathrm{out}$. Directly probing the transition $\ell_\mathrm{in}
\rightarrow \ell_\mathrm{out}$ poses a tremendous challenge, however,
and is in many cases even considered impossible. Herschbach \emph{et
al.} affectionately characterized the task of measuring and controlling
the impact parameter as the pursuit of the `forbidden fruit' in reaction
dynamics \cite{Herschbach2006,Anggara2018}. The number of partial waves
taking part in a collision event is directly connected to the particles'
de Broglie wavelength, and hence to the temperature or collision energy.
The higher the collision energy, the smaller the de Broglie wavelength,
and the larger the number of partial waves contributing to the
collision. These interfere with each other, which hampers a detailed
view on how the interaction potential transforms $\ell_\mathrm{in}$ into
$\ell_\mathrm{out}$ at the single partial wave level.

The main strategy to constrain the number of possible values for
$\ell_\mathrm{in}$ -- and thus to obtain a clear picture of the partial
wave evolution during a collision -- therefore lies with studying
collisions at the lowest possible temperature
\cite{Volz2005,McDonald:Nature535:122,Perreault:science358:356}. In the
extreme case where the temperature approaches zero kelvin one enters the
ultracold Wigner limit \cite{Wigner:PhysRev73:1002}. Here, the de
Broglie wavelength approaches infinity and we find a situation where in
principle only a single partial wave contributes -- for most systems
this is the $s$-wave --, unequivocally defining $\ell_\mathrm{in}$. At
slightly higher temperatures several incoming partial waves start to
contribute to the collision, but effects of single partial waves can
often still be traced through scattering resonances. Such a resonance
implies that the scattering partners briefly form a quasi-bound state,
and that only one -- or at most a few -- partial wave(s) dominate(s)
over other contributing waves. Tuning the collision energy to a
scattering resonance therefore offers the opportunity to `select' a
single partial wave from the initial conditions. This strategy works
best at energies just above the ultracold Wigner regime, where the
number of partial waves is still low such that scattering resonances
appear as prominent and distinct features in the cross sections.

Although spectacular progress has recently been made to experimentally
study molecular collisions at ever lower energies --- fully resolving
scattering resonances and even approaching the Wigner limit ---
\cite{Volz2005,McDonald:Nature535:122,Perreault:science358:356,Jankunas:JCP142:164305,Klein:NatPhys13:35,Lavert-Ofir:NatChem6:332,Bergeat:NatChem10:519,Bergeat:NatChem7:349,Bergeat:PRL125:143402,Jongh2020},
there are still caveats for studying the evolution of partial waves
during a collision event. In the ultracold regime the de Broglie
wavelength becomes so large that the topology of the underlying
interaction potential becomes irrelevant. Integral scattering cross
sections (ICSs) simply obey the universal Wigner laws, and the
differential cross sections (DCSs) show no intrinsic structure. In the
resonance regime, cross sections do possess intrinsic structure related
to the dominating partial wave, but so little energy is available that
the exit channel is necessarily governed by the same `selected' initial
partial wave and little is learnt about the \emph{evolution} of partial
waves \cite{Jongh2020}. In addition, typically only a very limited number of collision channels are energetically open, restricting the information that can be harvested from the potential accross its entire energy landscape.
Both effects pose a fundamental dilemma for
experiments designed to probe the collision dynamics at the
state-to-state partial wave level: The energy regime in which individual
partial waves are most easily probed, is also the energy regime in which
one becomes insensitive to how the interaction potential transforms
these waves during a collision.

Here we present a strategy that resolves this catch-22, and that allows
us to experimentally probe the transition $\ell_\mathrm{in} \rightarrow
\ell_\mathrm{out}$ in the cleanest way. To this end, we exploit a degree
of freedom of molecules that is foreign to atoms: rotational motion. The
key idea is to probe collisions in the translationally cold Wigner or
low-energy resonance regime such that the entrance channel is dominated
by a single partial wave only, and to add well-defined quanta of angular
momentum by rotationally exciting the molecules prior to the collision.
The study of rotational de-excitation collisions makes it possible to
trace how this additional amount of angular momentum is transferred into
translational motion. The outgoing de Broglie wavelength is
substantially decreased, which enriches the spectrum of outgoing partial
waves. The released kinetic energy has the additional experimental
benefit of increasing the otherwise minute collisional recoil energies,
making it easier to resolve structures in the DCS. Last but not least, the pre-collision excitation allows for a larger number of energetically allowed collision channels, offering more sensitive probes of the interaction potential.

Experimental studies on rotational de-excitation collisions at low
energies largely remain unexplored territory. In photodissociation of
ultracold molecules --- where an additional quantum of angular momentum
is offered by a photon --- similar ideas have previously been exploited
to probe partial wave dynamics \cite{McDonald:Nature535:122,Volz2005}.
For rotational state-to-state experiments such ideas were previously
explored theoretically \cite{Onvlee2016}, whereas recent measurements of
collisions involving rovibrationally excited NO molecules demonstrated
the possibility of monitoring rotational de-excitation in low-energy
collisions \cite{Amarasinghe2020}. However, the relatively high energies
involved in that study obscured clear resonance structures and the
partial wave fingerprints underlying the collisions.

We studied collisions between NO radicals and He atoms in a crossed
molecular beam machine. By optically preparing rotationally excited NO
radicals prior to the collision, we measured scattering resonances in
the ICSs and DCSs for rotational de-excitation collisions with energies
between 0.4 and 6 cm$^{-1}$ and an energy resolution up to 0.07
cm$^{-1}$. By tuning the collision energy to individual resonances, the
observed structures in the DCSs revealed how the additional quantum of
rotational energy transforms into the quantized relative motion of the
recoiling collision partners, revealing state-to-state energy transfer
at the full partial wave level. In addition, exploiting the principle of
detailed balance, we demonstrate that measurements of rotational
de-excitation processes provide a way to probe the time-reversed
inelastic excitation processes with much improved energy resolution.
Moreover, inelastic de-excitation processes do not have the energy
threshold of the corresponding excitation processes, which allows us to
measure resonance effects in cross sections at even lower energies. Our
results are in good agreement with state-of-the-art quantum chemistry calculations at the
CCSDT(Q) level.

\section*{Results and Discussion}

\subsection*{Low energy scattering resonances}

Low-energy rotational de-excitation collisions between He atoms and NO
molecules were studied in a crossed molecular beam machine that combines
Stark deceleration and velocity map imaging (VMI)
\cite{Meerakker:CR112:4828,Eppink:RSI68:3477}. The NO
molecules were state-selected and velocity controlled by the Stark
decelerator and subsequently excited into the $X\, ^2\Pi_{1/2}, \nu = 1,
j=3/2, f$ state by a quantum cascade laser (QCL). Here,
$X\,^2\Pi_{1/2}$, $\nu$ and $j$ indicate the electronic, vibrational and
rotational states, respectively and each rotational state is split into
two $\Lambda$-doublet components of opposite parity indicated by the
labels $e$ and $f$. In the following we will drop the electronic state
label $X\,^2\Pi_{1/2}$ and use the subscripts `in' and `out' to indicate
angular momentum quantum numbers for the reagents and products.
Rovibrational excitation of the NO radicals using the infrared radiation
offered by a QCL was chosen over a pure rotational transition (which
would require a THz source) for experimental reasons. The induced
vibrational excitation generally does not influence the scattering
process and the vibrational motion can in most cases be regarded a
spectator in the collision process (see Supplementary section 1.3).

\begin{figure}[!ht]
\centering
{\includegraphics[width=\textwidth]{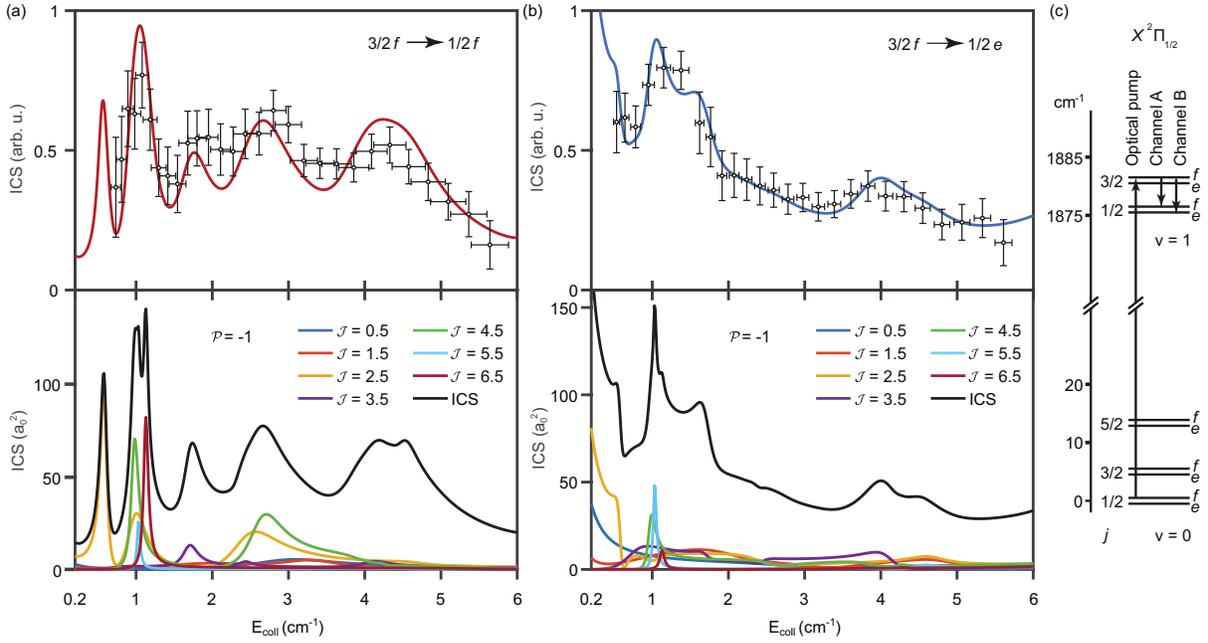}}
\caption{Collision energy dependence of the ICS and energy level diagram of the involved molecular states. Comparison between measured (data points with error bars) and calculated (solid curves) state-to-state inelastic scattering cross sections for the (a) $\nu = 1, j_\mathrm{in} = 3/2, f\rightarrow \nu = 1, j_\mathrm{out} = 1/2, f$ and (b) $\nu = 1, j_\mathrm{in} = 3/2, f \rightarrow \nu = 1, j_\mathrm{out} = 1/2, e$ transitions. Experimental data is given in arbitrary units (arb. u.). Vertical error bars display statistical uncertainties at $95\%$ of the confidence interval. Horizontal error bars represent the uncertainty in energy calibration. Solid lines show the theoretical curves convoluted with experimental energy spread (upper panels, overlay with data points), and individual partial wave contributions to the cross sections for $\mathcal{P} = -1$  (lower panels). (c) Energy level diagram of NO( $X\,^2\Pi_{1/2})$, indicating the optical transition used to prepare NO radicals in the $ \nu = 1, j = 3/2, f$ state (upward arrow), and the collision induced  rotational de-excitation transitions studied (downward arrows).  Energy level splittings due to $\Lambda$-doubling are greatly exaggerated for clarity.}
\label{fig:ICS}
\end{figure}

We measured state-to-state ICSs for the $\nu = 1, j_\mathrm{in} = 3/2,
f\rightarrow \nu = 1, j_\mathrm{out} = 1/2, f$ and $\nu = 1,
j_\mathrm{in} = 3/2, f\rightarrow \nu = 1, j_\mathrm{out} = 1/2, e$
rotational de-excitation channels of NO at energies between 0.4 and 6
cm$^{-1}$. We refer to these de-excitation transitions as channels A and
B, and present the results in Figures \ref{fig:ICS}a and \ref{fig:ICS}b,
respectively. An energy level diagram indicating the involved
rovibrational states and collision induced transitions is presented in
Figure \ref{fig:ICS}c. The experimentally obtained ICSs are compared
with theoretical predictions based on a state-of-the-art interaction
potential created with the coupled cluster method involving single,
double, triple and perturbative quadruple excitations [CCSDT(Q)], as
described in Ref. \cite{Jongh2020}. These theoretical curves were
convoluted with a Gaussian of varying width to account for the
experimental energy resolution, which ranged from 0.07 cm$^{-1}$ at the lowest energy to 0.4 cm$^{-1}$ at the highest energy. Experimental data points are vertically
scaled using root-mean-square fitting and corrected for density-to-flux
effects. For both channels, the measured ICS shows good agreement with
theoretical calculations, with peaks clearly indicating the presence of
partial wave resonances.

To investigate the nature of these resonances, we decomposed the
theoretically predicted cross sections in terms of the conserved total
angular momentum quantum number $\mathcal{J}$ and overall parity
$\mathcal{P}=\pm 1$. Figures 1a and 1b show the individual partial wave
contributions for $\mathcal{P}=-1$, which largely dominate the resonance
structures (see Supplementary section 2.1). The resonances are seen to
be relatively pure and are dominated by only a few values for
$\mathcal{J}$. The purity of the strong resonance feature in the $\nu =
1, j_\mathrm{in} = 3/2, f\rightarrow \nu = 1, j_\mathrm{out} = 1/2, f$
channel just below 1 cm$^{-1}$ is near-ideal, and is governed by only
one value for $\mathcal{J}$. Due to the conservation of $\mathcal{J}$
and $\mathcal{P}$, the values for the partial wave quantum numbers
$\ell_\mathrm{in}$ and $\ell_\mathrm{out}$ are constrained by the
initial $j_\mathrm{in}$ and final $j_\mathrm{out}$ rotational states of
NO --- taking the vector addition of angular momenta into account ---
via (see Supplementary section 2.2):
\begin{equation} \label{eq:totang}
	\vec{\mathcal{J}} = \vec{j}_\mathrm{in} + \vec{\ell}_\mathrm{in} = \vec{j}_\mathrm{out}  + \vec{\ell}_\mathrm{out},
\end{equation}
and
\begin{equation} \label{eq:totpar}
\mathcal{P} = \epsilon_\mathrm{in}(-1)^{j_\mathrm{in}-1/2+\ell_\mathrm{in}} = \epsilon_\mathrm{out}(-1)^{j_\mathrm{out}-1/2+\ell_\mathrm{out}},
\end{equation}
where $\epsilon$ equals $+1$ and $-1$ for rotational states with labels
$e$ and $f$, respectively.

Both rotational de-excitation transitions probed here ensure $\Delta j =
j_\mathrm{out} - j_\mathrm{in} = -1$, resulting in a richer evolution
from $\ell_\mathrm{in}$ to $\ell_\mathrm{out}$ compared to previous
low-energy NO-He experiments that were restricted by $\Delta j = 0$
\cite{Jongh2020}. Furthermore, they offer the additional benefit of a broader and more sensitive probe of the interaction potential. Channels A and B connect rotational levels with identical quantum numbers $j_{\mathrm{in}}$ and $j_\mathrm{out}$, but different values for $\epsilon$. Parity changing and parity conserving transitions are governed by different expansion terms of the interaction potential \cite{Aoiz:JPCA113:14636}, and their joint study yields a more complete picture of the interaction than measurements of the single $j_{\mathrm{in}}=1/2, f \rightarrow j_{\mathrm{out}}=1/2, e$  transition that is available without pre-excitation (see Supplementary section 2.4). For channels A and B, from application of equations
\ref{eq:totang} and \ref{eq:totpar} we can derive the propensity rules
for the partial wave quantum number $\Delta \ell = \pm 1$ and $\Delta
\ell = 0, \pm 2$, respectively, where $\Delta \ell > 0$ generally has
the largest contribution to the cross sections (see Supplementary section 2.3).

\begin{figure}[!ht]
\centering
{\includegraphics[width=\textwidth]{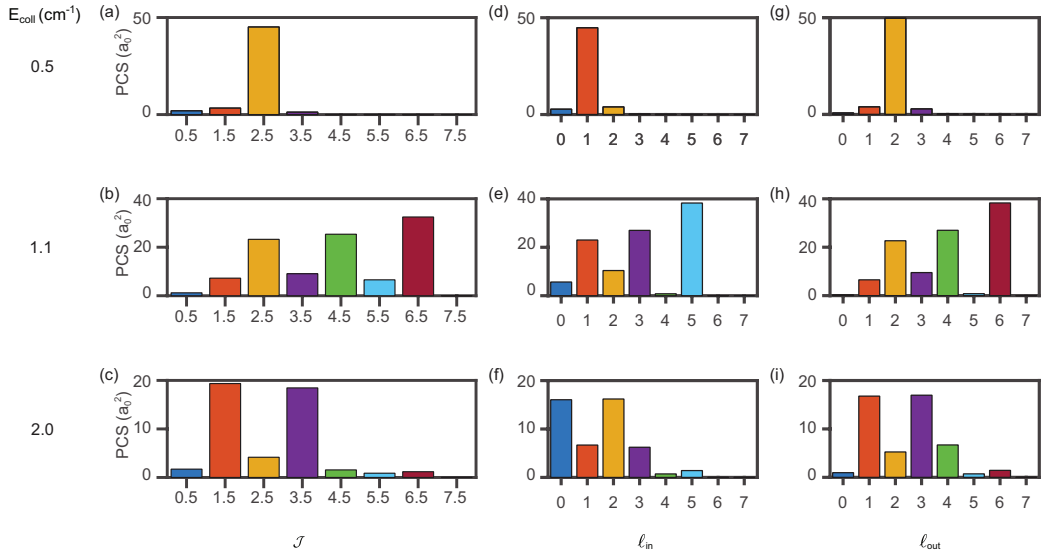}}
\caption{Theoretically predicted partial cross section (PCS) contributions for the total angular momentum $\mathcal{J}$ (a-c) as well as the incoming ($\ell_\mathrm{in}$) (d-f) and outgoing ($\ell_\mathrm{out}$) partial wave (g-i) for the $\nu = 1, j_\mathrm{in} = 3/2, f\rightarrow \nu = 1, j_\mathrm{out} = 1/2, f$ rotational de-excitation channel at three collision energies close to a scattering resonance. Experimental energy spreads ($\Delta$E$_\textrm{coll}$=0.08, 0.19, 0.28 cm$^{-1}$ (FWHM) for E$_{\textrm{coll}}$ = 0.5, 1.1, 2.0 cm$^{-1}$, respectively) have been taken into account.}
\label{fig:BarPartialWave}
\end{figure}

Figure \ref{fig:BarPartialWave} illustrates the relation between
$\mathcal{J}$, $\ell_\mathrm{in}$ and $\ell_\mathrm{out}$ for channel A
at collision energies of 0.5, 1.1, and 2.0 cm$^{-1}$, i.e., at energies
around the first three prominent scattering resonances occurring for
this transition. It is seen that de-excitation collisions add one unit
to the partial wave quantum number, consistent with the propensity rule
$\Delta \ell = 1$. At 0.5 cm$^{-1}$, a single total angular momentum
value dominates ($\mathcal{J} = 5/2$) causing the incoming $p$-waves
($\ell_\mathrm{in}=1$) to be almost fully converted into outgoing
$d$-waves ($\ell_\mathrm{out}=2$). This low energy resonance exquisitely
illustrates the advantage of measuring rotational de-excitation
processes to probe partial wave dynamics in collisions: The purity of
the incoming $p$-wave can only be achieved at energies approaching the
ultracold regime, while the introduction of outgoing partial waves with
higher $\ell$ is only allowed because of the released quantum of angular
momentum stored in the rotation of the incoming NO molecule. A similar
analysis was made for the resonances observed in channel B (see
Supplementary Information), where the evolution of partial waves was
found to follow the propensity rule $\Delta \ell = 0$ or $2$.

\subsection*{Evolution of partial waves}

The evolution of partial waves $\ell_\mathrm{in} \rightarrow
\ell_\mathrm{out}$ in a collision event is directly encoded in the
measured DCS, which reflects the outgoing partial waves, each
corresponding to a spherical harmonic of degree $\ell_{\mathrm{out}}$. The number of nodes in the DCS is related
with the largest value of $\ell_\mathrm{out}$ \cite{NIST:Handbook:2010},
leading to an isotropic pattern in the case of pure $s$-wave scattering,
a pattern with at most a single node for pure $p$-wave scattering, and
so on. We therefore probed DCSs for both collision induced $\Delta j =
-1$ de-excitation transitions by measuring velocity mapped ion images at
several selected collision energies between 0.4 and 2.0 cm$^{-1}$. To
elucidate the effect of the additional quantum of angular momentum
released in the de-excitation collision, we compared our results with
measurements from $\Delta j=0$ transitions. To this end, we measured ion
images for the previously \cite{Jongh2020} investigated $\nu = 0,
j_\mathrm{in} = 1/2, f \rightarrow \nu = 0, j_\mathrm{out} = 1/2, e$
parity de-excitation channel at the same collision energies. Effects of
the difference in vibrational state exist but in most cases were small
enough to warrant direct comparison and trace the effects of the
released quantum of angular momentum in the rotational de-excitation
channels (Supplementary section 1.3).

\begin{figure}
\centering
{\includegraphics[width=\textwidth]{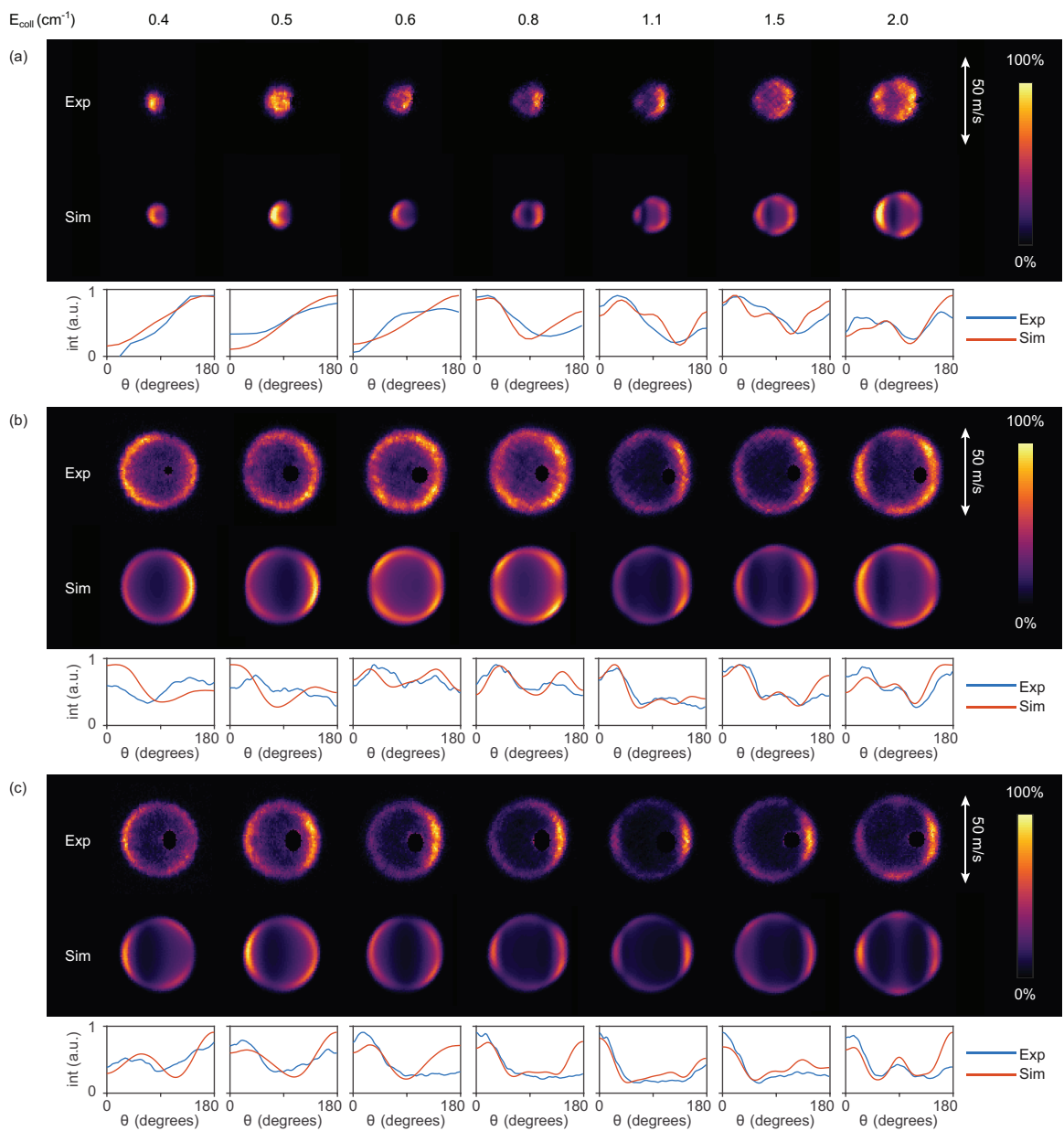}}
\caption{Experimental and simulated ion images at selected collision energies for the (a) $\nu = 0, j_\mathrm{in} = 1/2$ $f \rightarrow \nu = 0, j_\mathrm{out} = 1/2$ $e$ parity de-excitation collisions, (b) $\nu = 1, j_\mathrm{in} = 3/2,$ $f \rightarrow \nu = 1, j_\mathrm{out} = 1/2,$ $e$ rotational de-excitation channel, and the (c) $\nu = 1, j_\mathrm{in} = 3/2,$ $f \rightarrow \nu = 1, j_\mathrm{out} = 1/2,$ $f$ rotational de-excitation channel. Collision energies are indicated in the top of the figure and are identical throughout each column. Experimental (Exp) and simulated (Sim) ion images are rotated such that the relative velocity vector is oriented horizontally, with the forward scattering direction on the right side of each image. Color scale represents intensity and the overall intensity scale for the simulated images is fitted to that of the experiment. A small portion of the images is masked due to imperfect state selection of the incoming NO molecules. Angular distributions are extracted from the ion images and plotted for each channel and collision energy.}
\label{fig:DCS1}
\end{figure}

The experimentally obtained ion images are shown in Figure
\ref{fig:DCS1} for each scattering channel and collision energy. In
these images, the angular distribution reflects the
underlying DCS. Indeed, the distributions for $\Delta
j =-1$ rotational de-excitation transitions show more complex features
with additional nodes compared to the $\Delta j = 0$ transition, a
direct consequence of the transition of rotational into translational
angular momentum. The rotational de-excitation channels furthermore lead
to distinctly larger ion images than the $\Delta j = 0$ parity
de-excitation channel as a result of the transfer of internal into
kinetic energy. This enables a higher resolution in the measured angular
distributions of the scattering products. In addition, the scattering
signal is well separated from the image position of the reagent packet,
referred to as the beam spot, which is caused by imperfect state
selection of the reagent beam and is masked in the images by a black
spot. Simulations of the expected images based on DCSs computed with the
CCSDT(Q) potential are shown as well, together with the angular
distributions extracted from both the experimental and simulated images.
These simulations take the full kinematics of the experiment into
account, such as collision energy spreads and detection efficiencies.
Overall, good agreement is obtained between theory and experiment,
although for channel A disparities are found in the backward scattering
region ($\theta \approx 180^\circ$) at 0.6 and 0.8 cm$^{-1}$. We speculate that this may be due to the strongly varying DCS across this energy range near the steep rising edge of a resonance.  

The increased complexity of the DCS structure caused by the added
quantum of angular momentum can be directly related to the evolution of
partial waves during the collision. This can be seen most clearly at the
lowest energies probed. Here, the $\Delta j =0$ channel displays a
rather simple angular distribution, consistent with the interference of
only $s$- and $p$-wave distributions found at these energies
\cite{Jongh2020}. By contrast, the $\Delta j = 1$ rotational
de-excitation transitions measured at the same collision energy lead to
structures with a larger number of nodes, indicative for the additional
contributions of outgoing waves with
$\ell_\mathrm{out}>\ell_\mathrm{in}$ and consistent with the propensity
rules for the partial waves discussed above. Specifically, for channel A
at $E_\mathrm{coll} = 0.5$ cm$^{-1}$ --- which exhibits the most
pristine partial wave resonance probed here --- the observed two-node
structure in the angular distribution directly reveals the evolution of
a $p$- wave into a $d$-wave that underlies the scattering event.

\subsection*{Detailed balance}

The DCS measurements presented in Figure \ref{fig:DCS1} thus display
many advantages introduced by adding an internal quantum of angular
momentum before the collision. Yet there is another aspect that can be
expoited to study low-energy collision phenomena: owing to the principle
of time-reversal invariance, de-excitation collisions probe the same
partial wave dynamics that underlies the corresponding excitation
process, but with a significantly higher experimental energy resolution.
This principle --- often also referred to as the principle of
microscopic detailed balance \cite{Boltzmann1872,Einstein1916} --- dictates that the DCS for a scattering
event is identical to the DCS for the time-reversed process apart from a
constant factor, with time-reversal effectively swapping
$(j_\mathrm{in},\ell_\mathrm{in})$ and
$(j_\mathrm{out},\ell_\mathrm{out})$ (see Supplementary
sections 2.5 and 2.6), provided that the total (kinetic + rotational) energy
remains the same. Whereas the excitation process has an energetic
threshold corresponding to the internal energy of the excited state,
de-excitation processes can be probed at arbitrarily low energies where
the experimental energy resolution is typically much higher.

The experimental arrangement used here, which offers access to
well-resolved scattering resonances and the possibility to prepare
molecules in a rotationally excited state, yields the unique opportunity
to probe two complementary time-reversed processes under similar
circumstances. To illustrate this, we compared the scattering images for
the $\nu = 1, j_\mathrm{in} = 3/2,$ $f \rightarrow \nu=1, j_\mathrm{out}
= 1/2,$ $f$ de-excitation channel at collision energies
$E_\mathrm{coll}$ (from Figure \ref{fig:DCS1}) with images taken for the
$\nu=0, j_\mathrm{in} = 1/2, f \rightarrow \nu = 0, j_\mathrm{out} =
3/2, f$ excitation transition at energies $E_\mathrm{coll} + \Delta
E_\mathrm{rot}$ , where $\Delta E_\mathrm{rot} = 5.02$ cm$^{-1}$ is the
energy difference between the $j=1/2$ and $j=3/2$ rotational states of
NO. Under these conditions, we can regard these transitions as each
other's time-reversed process for which the reversibility principle
should hold, provided that the vibrational state is a spectator in the
collision process. From theoretical calculations, we found that in most
cases the effects from the difference in vibrational states can indeed
be neglected (see Supplementary section 1.3).

\begin{figure}[!ht]
\centering
{\includegraphics[width=.99\textwidth]{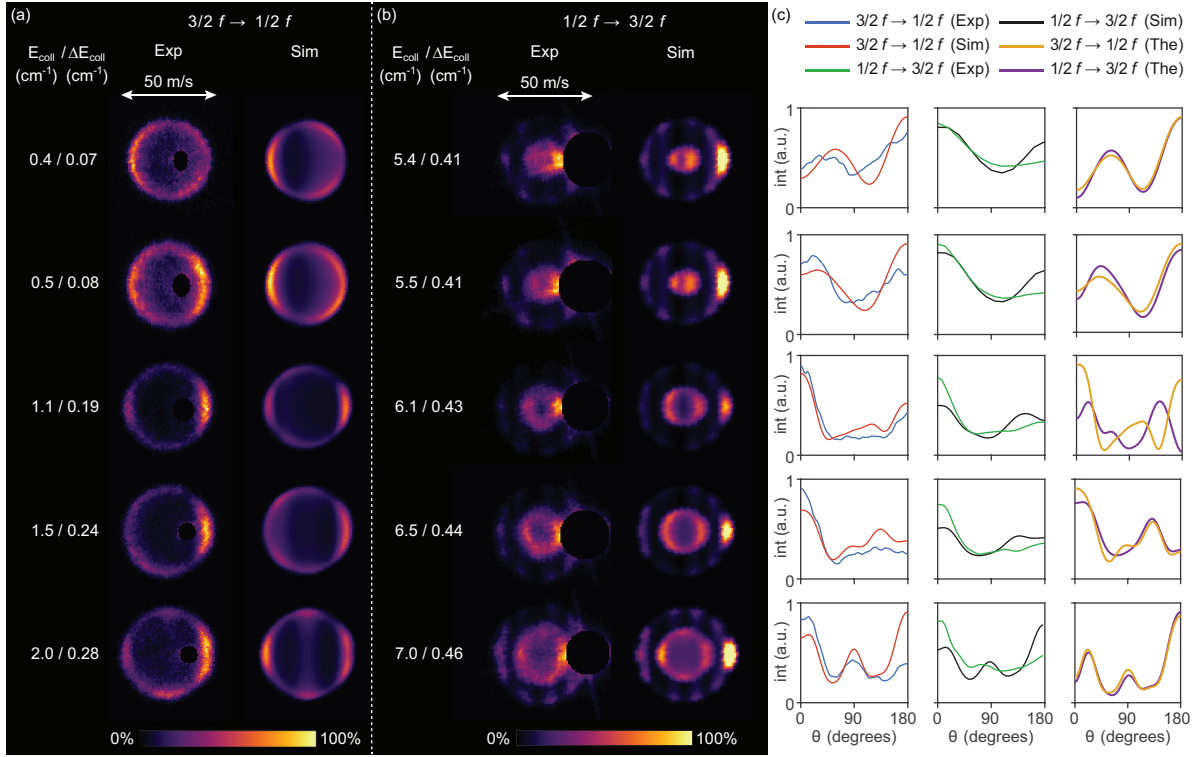}}
\caption{Experimental (Exp) and simulated (Sim) ion images for two inelastic scattering processes subject to the detailed balance principle. (a) Images for the $\nu = 1, j_\mathrm{in} = 3/2,$ $f \rightarrow \nu=1, j_\mathrm{out} = 1/2,$ $f$ rotational de-excitation process at selected collision energies $E_\mathrm{coll}$. (b) Images for the $\nu = 0, j_\mathrm{in} = 1/2,$ $f \rightarrow \nu = 0, j_\mathrm{out}= 3/2,$ $f$ rotational excitation channel at energies $E_\mathrm{col} + \Delta E_\mathrm{rot}$, with $\Delta E_\mathrm{rot} = 5 $ cm$^{-1}$. The experimental collision energy spread $\Delta$E$_\textrm{coll}$ is indicated for each image. An outer ring is visible that is caused by elastic scattering of NO molecules in the $\nu = 0, j = 3/2,$ $f$ state. Part of the forward scattering direction of these outer rings are masked due to imperfect state selection of the NO radicals. Images are oriented such that the relative velocity vector is aligned horizontally, with the forward direction on the right.  (c) Extracted angular distributions for the inelastic channels plotted for each collision energy as well as normalized purely theoretical DCSs (The) within the respective vibrational states.}
\label{fig:DCS_DB}
\end{figure}

The images pertaining to the $j_\mathrm{in}=3/2 \rightarrow
j_\mathrm{out}=1/2$ de-excitation channel from the last row of Figure
\ref{fig:DCS1} are re-plotted for selected energies in Figure
\ref{fig:DCS_DB}, together with the obtained ion images for the
corresponding $j_\mathrm{in}=1/2 \rightarrow j_\mathrm{out}=3/2$
excitation channel. For both channels, the simulated images and
extracted angular distributions are shown as well. The ion images for
the $j_\mathrm{in}=1/2 \rightarrow j_\mathrm{out}=3/2$ excitation
channel show two concentric rings, with the smaller ring pertaining to
the $\nu = 0, j_\mathrm{in} = 1/2, f \rightarrow \nu = 0, j_\mathrm{out}
= 3/2, f$ transition and the larger ring caused by elastic collisions
involving NO radicals in the $\nu = 0, j = 3/2, f$ state, which is
present in minute amounts in the primary NO packet. This outer ring
appears as a relatively strong feature due to the much larger cross
section for elastic scattering compared to inelastic collisions.

Overall, good agreement is obtained between the experimental and the
simulated images. However, at the lower energies probed the angular
distributions of the de-excitation and its time-reversed excitation
process appear rather different, seemingly violating the principle of
detailed balance. Yet, the purely theoretical DCSs (within the
respective vibrational states) underlying these simulated images are
shown in the rightmost column of Figure \ref{fig:DCS_DB}, and are found
to be indeed identical, except at a collision energy of 1.1 cm$^{-1}$
(6.1 cm$^{-1}$ for the excitation channel) in the vicinity of a
scattering resonance, where the vibration causes a notable
difference in the cross sections (Supplementary section 1.3). This
apparent contradiction is merely a direct consequence of the kinematics
of the experiment: The higher collision energy for the excitation
process results in a larger collision energy spread, and the transfer of
kinetic into rotational energy results in a smaller image. Both have
substantial consequences for the observable angular distribution in a
scattering image: The larger energy spread causes averaging of the cross
sections within a broader energy range, while structures in the DCS are
more easily blurred in images with small diameter. These effects are taken into account in the simulations,
and cause the angular distributions of the simulated scattering images
to deviate, even though the underlying DCSs obey detailed balance.

These results further underline that by virtue of time-reversal
invariance, rotational excitation processes are actually probed with
superior energy and angular resolution by measuring the corresponding
de-excitation channel instead. For the energy ranges investigated here,
the collision energy spread (FWHM) for the de-excitation collisions
ranges from 0.07 to 0.3 cm$^{-1}$ whereas it ranges from 0.41 to 0.44
cm$^{-1}$ for the excitation collisions. At the lowest energies probed,
this enhancement in resolution can thus reach up to a factor 5 or
higher. Furthermore, the larger image diameter afforded by the
de-excitation measurements translates into an angular resolution of
approximately 3$^\circ$, whereas the angular resolution of the
excitation images is between 5$^\circ$-18$^\circ$.

\section*{Conclusion}

The study of low-energy molecular collisions provides unique
opportunities to trace the partial wave dynamics underlying a scattering
event. The limited set of incoming partial waves available at these
energies is a prerequisite for the mapping of their dynamics, but the
typically equally limited set of outgoing partial waves reduces the
sensitivity to the interaction potential. Using rovibrational excitation
of the reagent molecules, and subsequently studying de-excitation
inelastic scattering, we were able to resolve this catch-22 as it
introduces an additional available quantum of angular momentum to the
outgoing spectrum of partial waves. We traced its effects on the
scattering dynamics by measuring integral and differential cross
sections at energies in the vicinity of scattering resonances, revealing
the evolution of partial waves throughout the scattering process. 
Furthermore, the larger number of energetically accessible inelastic collision channels afforded by the pre-collision excitation offered more detailed probes of the interaction potential accross its entire energy landscape.
Rotational de-excitation collisions were also demonstrated to offer
superior resolution --- in both the collision energy and the angular
distribution of the scattering products --- compared to their
time-reversed rotational excitation counterparts.  Optical excitation of reagent molecules to any
initial rotational state can in principle be achieved by sequential
rotational excitations, offering the unique opportunity to study
how controlled amounts of angular momentum stored in translationally
ultracold molecules is released at the full partial wave level.

\section*{Acknowledgements}
This work is part of the research program of the Netherlands Organization for Scientific Research (NWO). S.Y.T.v.d.M. acknowledges support from the European Research Council (ERC) under the European Union’s Seventh Framework Program (FP7/2007-2013/ERC Grant Agreement
No. 335646 MOLBIL) and from the ERC under the European Union’s Horizon 2020 Research and Innovation Program (Grant Agreement No. 817947 FICOMOL).
We thank N. Janssen and A. van Roij for expert technical support. We thank F.J.M. Harren for stimulating discussions regarding the optical excitation of NO using a quantum cascade laser. We thank Tijs Karman for fruitful discussions and for critically reading the manuscript.

\section*{Author Contributions}
The project was conceived by S.Y.T.v.d.M. The experiments were carried out by T.d.J., Q.S. and S.K.. Methods to rovibrationally excite NO using a quantum cascade laser were developed by G.A. and Q.S.. Data analysis and simulations were performed by T.d.J. Theoretical calculations were performed by M.B., A.v.d.A. and G.C.G. The manuscript was written by T.d.J. and S.Y.T.v.d.M. with contributions from all authors. All authors were involved in the interpretation of the data and the preparation of the manuscript.

\section*{Competing Interest Statement}
There are no competing interests.

\section*{Methods}
The experiments were performed in a crossed-molecular beam apparatus described previously \cite{Jongh2020methods}. A supersonic beam of 5\% NO seeded in Kr or Ar was expanded through a Nijmegen Pulsed Valve \cite{Yan:RSI84:023102} and subsequently entered a 2.6 m long Stark decelerator. The resulting NO molecules in the $X ^2\Pi_{1/2}, \nu = 0, j = 1/2, f$ state were velocity selected with a mean velocity between 480 and 720 m/s and a longitudinal spread of approximately 2.1 m/s $(1 \sigma)$. At a distance of 5 mm from the exit of the Stark decelerator these molecules were rovibrationally excited by a distributed feedback QCL (Thorlabs, QD5316CM) at a wavelength of approximately 5.3 $\mu$m, inducing the $\nu' = 1, j' = 3/2, f, F' = 5/2 \leftarrow \nu'' = 0, j'' = 1/2, f, F'' = 3/2$ transition in order to obtain the highest yield of excited NO radicals (see Supplementary section 1.2). Here $F$ indicates the initial and final total angular momentum including nuclear spin. The nuclear spin can be treated as a spectator as well \cite{Ball1999}, which was verfied by studying rotational inelastic collisions when inducing other hyperfine transitions (see Supplementary section 1.2).

The packet of NO radicals intercepted a neat beam of He under an angle of 5$^\circ$. This secondary beam was expanded through a commercially available Even-Lavie Valve \cite{Even:AiC2014:636042} which was cooled by a cryogenic cold head to temperatures between 15 K and 30 K, resulting in He velocities between 460 and 575 m/s. Calibration of the He velocity and the energy resolution was performed by measuring a previously identified resonance peak in the $\nu = 0, j_\mathrm{in} = 1/2, f \rightarrow \nu = 0, j_\mathrm{out} = 1/2, e$ channel of NO-He collisions near a collision energy of 1 cm$^{-1}$ \cite{Jongh2020methods}. Scattered NO radicals were state-selectively detected using 1+1' resonance enhanced multiphoton ionization involving the (1-1) band of the $A^2\Sigma^+ \leftarrow X ^2\Pi_{1/2}$ transition using two dye lasers pumped by a single Nd:YAG laser. The ions were then mapped onto a microchannel plate using an advanced velocity map imaging spectrometer yielding a resolution of $0.8$ m/s/pixel \cite{Plomp2020}.

For DCS measurements, a small image correction --- a 5 \% compression of the image in the vertical direction --- was made to mitigate the effect of slight distortions expected to be caused by the presence of external magnetic fields. ICS measurements were performed with the imaging spectrometer out of focus in order to avoid detector saturation. The collision energy was tuned by scanning the NO velocity between 580 and 720 m/s using the Stark decelerator. In order to obtain optimal resolution \cite{Scharfenberg2011}, we measured two energy ranges separately using different He velocities (approximately 480 and 550 m/s). Results were corrected for beam density and flux-to-density effects. Details on the simulations are given elsewhere \cite{Zastrow:NatChem6:216}.\\

\end{document}